\documentclass[twocolumn,pre,twoside,floatfix]{revtex4-1}
\usepackage{graphics,psfrag}
\usepackage{graphicx,psfrag}
\usepackage{amsmath}
\usepackage{amssymb}
\usepackage{xfrac}
\usepackage{ifsym}
\usepackage{textcomp}
\usepackage{color}
\usepackage{ulem}
\newcommand{\be}{\begin{equation}}
\newcommand{\ee}{\end{equation}}

\newcommand{\Tw}{$T_{\text w}$}
\newcommand{\Tws}{$T^*_{\text w}$}
\begin{document}


\title{Wetting in electrolyte solutions}

\author{Ingrid Ibagon}
\email{ingrid@is.mpg.de}
\author{Markus Bier}
\email{bier@is.mpg.de}
\author{S.\ Dietrich}
\affiliation
{
   Max-Planck-Institut f\"ur Intelligente Systeme, 
   Heisenbergstr.\ 3,
   70569 Stuttgart,
   Germany, 
   and
   IV. Institut f\"ur Theoretische Physik,
   Universit\"at Stuttgart,
   Pfaffenwaldring 57,
   70569 Stuttgart,
   Germany
}

\date{\today}

\begin{abstract}
   Wetting of a charged substrate by an electrolyte solution is investigated
by means of classical density functional theory applied to a lattice model. Within the present
model the pure, i.e., salt-free solvent, for which all interactions are of the
nearest-neighbor type only, exhibits a second-order wetting transition for all strengths of the
substrate-particle and the particle-particle interactions
 for which the wetting transition
temperature
is nonzero. The influences of the
substrate charge density and of the ionic strength on the wetting
transition temperature and on the order of the wetting
transition are studied. If the substrate is neutral, the addition of salt to the solvent
changes neither the order nor the transition temperature of the wetting transition of the system.
If the
surface charge is nonzero, upon adding salt this continuous wetting transition changes to
first-order within the wide range of substrate surface charge densities and ionic strengths studied
here. As the
substrate
surface charge density is increased, at fixed ionic strength, the wetting transition temperature
decreases
and the
prewetting line associated with the first-order wetting transition becomes longer. This decrease of
the wetting transition temperature upon increasing the surface charge density becomes more
pronounced by
decreasing the
ionic strength.
\end{abstract}

\maketitle


\section{Introduction}

Wetting transitions are surface phase transitions which occur whenever a phase C intrudes at
the interface between two phases A and B, with either A, B, and C in thermodynamic
coexistence or with A as a spectator phase and B and C in thermodynamic coexistence. As an example,
A
is an inert substrate and B and C are the gas and the liquid phase, respectively, of a simple
fluid. The thickness of the intruding liquid film can be either finite (incomplete wetting) or
macroscopically large (complete wetting) upon approaching  gas-liquid
coexistence along an isotherm. The transition at  two-phase coexistence from incomplete to
complete wetting
occurs at the wetting transition temperature $T=T_{\text w}$. It
can be either continuous (second-order), in which case the film thickness diverges smoothly as
$T\to T_{\text w}$ along two-phase coexistence, or discontinuous (first-order), implying a
macroscopically
large
jump of the film thickness 
from a finite value below $T_{\text w}$ to a macroscopically large one above $T_{\text w}$. In the
surface
phase diagram a first-order wetting 
transition has a prewetting line associated with it which is connected tangentially to the
gas-liquid coexistence
line at \Tw, extends into the gas phase region, and ends at a critical point.
The richness of wetting phenomena has been covered by various reviews \cite{deGennes, Dietrich1988,
Schick, Bonn2001, Bonn2009}.

So far, to a large extent, wetting studies have been devoted to fluids composed of electrically
neutral molecules. However,
for numerous real systems the presence of ions is either of
crucial importance for wetting phenomena, such as electrowetting \cite{Mugele}, or
unavoidable
because many substrates release ions once they are brought into contact with polar solvents
\cite{Everett}. For example, electrowetting refers to the change of the substrate-fluid interfacial
tension as a response to an applied electrostatic potential difference between the substrate and
the fluid bulk. This effect offers numerous applications in devices based on the manipulation
of tiny
amounts of liquids, such as microfluidic devices \cite{pollack2000, Valev2003}.
Theoretical studies of those systems started back in 1938 when Langmuir developed a
model to determine the equilibrium thickness of water layers on planar surfaces in contact with
undersaturated water vapor, based on the calculation of the repulsive force between two plates
immersed in electrolyte solutions \cite{Langmuir1938}. The typical values for the equilibrium layer
thickness as predicted by Langmuir's formula were confirmed experimentally \cite{Hall1970} and the
experimental data were used to analyze the effect of various contributions to the disjoining
pressure onto the stability of the wetting films \cite{Derjaguin1974}. Some years later Kayser
generalized Langmuir's model for the equilibrium thickness of wetting layers to liquid mixtures of
polar and non-polar components in contact with ionizables substrates \cite{Kayser1986}; in 
contact  with the wetting liquid these substrates donate ions to the liquid which act as
counterions to the emerging opposite charge left on the substrate with overall charge neutrality.
This analysis was followed up  by
including the effect of added salt the ions of which do not stem from the
substrate \cite{Kayser1988}. These papers did not
address the issue of wetting transitions at coexistence but rather focused on the thickness of
the wetting
layer and the behavior of the disjoining pressure. For wetting films of solvents without added
salt, i.e., with counterions only, Langmuir \cite{Langmuir1938} and Kayser \cite{Kayser1986} found
that
the film thickness $l$ increases as $l\sim(\Delta \mu)^{-1/2}$, with $\Delta
\mu=\mu_{co}-\mu$, as the
chemical potential $\mu$ approaches its value $\mu_{co}$ at coexistence from the vapor side
($\mu<\mu_{co}$). In contrast, wetting films without ions and at neutral substrates but with
van der Waals
interactions (which are not taken into account in our model) lead to $l\sim(\Delta
\mu)^{-1/4}$ or
$l\sim(\Delta \mu)^{-1/3}$, depending on whether retardation effects are taken into account or not,
respectively
\cite{Dietrich1988}.
In the case that the effect of added salt dominates van der Waals interactions Kayser
\cite{Kayser1988} found $l\sim\ln(\Delta \mu)$ as it holds for short-ranged interactions. 

Only recently theoretical investigations
concerning wetting transitions of electrolyte solutions at charged solid substrates have emerged
\cite{Denesyuk2004, Oleksy2009, Oleksy2010}. In
Ref. \cite{Denesyuk2004} the effect of adding ions onto the wetting behavior of the pure
solvent was studied by using Cahn's phenomenological theory \cite{deGennes, Dietrich1988,
Schick, Bonn2001} for the solvent combined with the
Poisson-Boltzmann theory for the ions. This model does not take
into account the solvent particles explicitly, neglecting the coupling between solvent particles and
ions. On the other hand, the model in Ref. \cite{Oleksy2009} takes all three types of
particles (solvent, cations, and anions) explicitly into account in terms of hard spheres of
different diameters with a Yukawa
attraction between all pairs and the Coulomb interaction between ions. The model was studied by
using Rosenfeld's density
functional theory \cite{Rosenfeld1988, Rosenfeld1989} combined with a mean-field approximation for
the Yukawa and the electrostatic interactions. Within this model, the polar nature of the solvent
molecules was ignored; it was included in a subsequent article by the same authors in which
the solvent particles were represented by dipolar hard spheres \cite{Oleksy2010}. However,
for technical reasons, the numerical analyses of these
continuum models in which all three types of particles are treated explicitly on a
microscopic
level were limited to small system sizes. Therefore Refs. \cite{Oleksy2009, Oleksy2010} focused on
the case of strong screening of the Coulomb interactions
which is provided by large ionic strengths, i.e., large ion concentrations. However, the approaches
used in Refs. \cite{Denesyuk2004, Oleksy2009, Oleksy2010} are not reliable for large ionic
strengths due to the use of Poisson-Boltzmann theory for the electrostatic interactions which has
been proved to be valid only for low ionic concentrations \cite{Torrie1979}.

In order to overcome
these problems
we introduce a lattice model for an electrolyte exposed to a charged substrate which takes into
account all three components via density functional theory and offers the
possibility to study significantly broader interfacial regions. In Sec. \ref{MF} we introduce this
model and
the approximate density functional. In Sec. \ref{res} we present our results for the bulk
properties and
the wetting
phenomena for both the 
salt-free solvent and the electrolyte
solution. We conclude and
summarize our main results in Sec. \ref{CS}.

\section{Model and density functional theory}\label{MF}
\subsection{Model}
We study a lattice model for an electrolyte solution in contact with a charged
wall.  The
solution consists of three components: solvent $(0)$, anions
$(-)$, and cations $(+)$. The coordinate perpendicular to the wall is $z$. The region
above the wall, accessible
to the electrolyte
components, 
is divided into a set of cells the centers of which form a simple cubic lattice $\{\bf r\}$
with
lattice constant
$a$. The volume $a^3$ of a cell corresponds roughly to the volumes of the particles, which are
assumed to be of similar size. The centers of the molecules in the top layer of
the substrate form the
plane $z=0$. At closest approach the centers of the solvent molecules and ions are at $z=a$. The
plane
$z=a/2$ is taken to be the surface of the planar wall. 
Each cell is either empty or occupied by a single particle. This mimics the steric hard core
repulsion between all particles. Particles at different
sites
interact among each other via
an attractive nearest-neighbor interaction of strength $u$ which is taken to be
the same for all
pairs of particles. In addition, ion pairs interact via the Coulomb potential. The solvent
particles are taken to carry a dipole moment.
 
The wall attracts particles only in the first adjacent layer via an interaction potential of
strength
$u_{\text w}$ which is the same for all species. In addition it can carry a homogeneous surface
charge density $\tilde\sigma=\sigma ea^{-2}$ which is taken to be localized in the plane
$z=a/2$ and which 
interacts
electrostatically with the ions; $e>0$ is the elementary charge. Since we focus on the influence of
the ions onto wetting phenomena we
refrain from considering
the more realistic, long-ranged van der Waals forces which are known to be relevant for wetting
transitions \cite{Dietrich1988}. Within the mean-field theory we shall use, the choice of
nearest-neighbor interactions provides a significant computational bonus which we want to exploit
in favor of our core concern stated above. 

The corresponding lattice-gas Hamiltonian for this system reads
\be\label{h1}
\begin{split}
H&=\frac12 \sum_{\substack{{\bf r},{\bf r'}\\ {\bf r}\neq {\bf
r'}}}\sum_{i,j}n_i({\bf r})n_j({\bf r'})w\left(|{\bf r}-{\bf
r'}|\right)
\\&+\frac 12\sum_{\substack{{\bf r},{\bf
r'}\\ {\bf r}\neq{\bf
r'}}}\sum_{i,j}\frac{e^2q_iq_jn_i({\bf
r})n_j(\bf {r\,'})}{4\pi\varepsilon_0|{\bf r}-{\bf r'}|}
\\&+\sum_{\substack{{\bf r},{\bf
r'}\\ {\bf r}\neq{\bf
r'}}}\sum_{i,j}\frac{eq_in_i({\bf
r}){\bf m_j({\bf r'})\cdot({\bf r}-{\bf r'})}}{4\pi\varepsilon_0|{\bf r}-{\bf
r'}|^3}
\\&+\frac12\sum_{\substack{{\bf r},{\bf
r'}\\ {\bf r}\neq{\bf
r'}}}\sum_{i,j}\left[\frac{{\bf m}_i({\bf
r})\cdot{\bf m}_j({\bf r'})}{4\pi\varepsilon_0|{\bf r}-{\bf r'}|^3}\right.
\\&\left.-\frac{3
\left({\bf m}_i({\bf r})\cdot ({\bf r}-{\bf r'})\right)\left({\bf m}_j({\bf r'})\cdot ({\bf r}-{\bf
r'})\right)}{4\pi\varepsilon_0|{\bf
r}-{\bf r'}|^5}\right]
\\&-\sum_{{\bf r}}\sum_iu_{\text w}\delta_{z, a}n_i({\bf r})-
\frac{\tilde\sigma}{2\varepsilon_0}\sum_{{\bf r}}\sum_iq_in_i({\bf
r})z
\\&-\frac{\tilde\sigma}{2\varepsilon_0}\sum_{{\bf r}}\sum_i{\bf m}_i({\bf r})\cdot \hat {\bf e}_z
\end{split}
\ee 
where $n_i({\bf r})$ are occupation number variables, which are either 0 or 1 according to whether
the cell at the discrete position ${\bf r}=({\bf r_{||}},z\ge a)=(x,y,z\ge a)=(ma,na,pa)$ with
$m,n\in\mathbb{Z}$, $|m|\le\bar M/2$ and $|n|\le\bar N/2$, and $p=1,2,3,\cdots,\bar L$ is
empty or occupied by a particle (there is no double occupancy); $i,j=0,+,-$,
$eq_i$ is the particle charge with $q_0=0$ and $q_{\pm}=\pm1$; ${\bf m_i}({\bf r})$ is the
particle
dipole
moment at ${\bf r}$ (we consider the typical situation of a polar solvent
and of ions without permanent electric dipoles, i.e., ${\bf m_\pm}=0$); $w\left(|{\bf r}-{\bf
r'}|\right)=-u$ for nearest neighbors ($u>0$ corresponds to attraction) and
$w\left(|{\bf r}-{\bf r'}|\right)=0$ beyond;  $-u_{\text w}$ is the strength of the
attractive ($u_{\text w}>0$) substrate potential acting on the first
layer $z=a$. For the charge density $\tilde \rho({\bf r})=\tilde \sigma\delta(z-a/2)$ on
a
substrate with radial extension $R_0$ the electrostatic potential is given by $\tilde\phi({\bf
r})=\int{ \!d^3r'\!\tfrac{\tilde \rho({\bf r'})}{4\pi\varepsilon_0|{\bf r}-{\bf r'}|}}=\tfrac{\tilde
\sigma}{2\varepsilon_0}(\sqrt{R_0^2+(z-a/2)^2}-|z-a/2|)\to-\tfrac{\tilde
\sigma}{2\varepsilon_0}z+const.$ for $R_0\gg|z-a/2|$ and $z>a/2$. In this regime of being close to
the charged wall the electric field is uniform \cite{Jackson}. Therefore the actual position of the
charged wall enters the
electrostatic potential, and thus the Hamiltonian in Eq. (\ref{h1}), only via an irrelevant
additive constant. The potential energy of a
dipole
moment ${\bf m}_i({\bf r})$ in the electric field $\tilde {\bf E}({\bf
r})=\tilde\nabla\tilde\phi({\bf r})\to\tfrac{\tilde\sigma}{2\varepsilon_0}\hat{\bf e}_z=
const.$ of the
surface charge is given by $-{\bf m}_i({\bf r})\cdot\tilde {\bf E}$. In Eq. (\ref{h1}) we consider
only
charge neutral configurations $\{n_i({\bf r})\}$, i.e., $\sum\limits_{\bf r}\left(n_+({\bf
r})-n_-({\bf
r})\right)=-\bar M\bar N\sigma$ with ${\bf r}\in V=\bar M\bar N\bar La^3$.

For weak external electric fields the polarization is expected to exhibit a linear
response behavior
\cite{Jackson}. In this case, it has been
shown that the relative permittivity  $\varepsilon$ of microscopic models like the one in
Eq. (\ref{h1})
 can be expressed in terms of molecular properties such as the dipole moment and the
polarizability \cite{Wertheim, Carnie}.
In order to simplify our model, the polar nature of the solvent is taken into account effectively
via the
relative permittivity $\varepsilon$ of
the electrolyte solution which is assumed to depend on the solvent configuration $n_0({\bf
r})$ but not on the configuration
of the ions $n_\pm({\bf r})$ because the orientational polarization, i.e., the polarization
due
to the permanent dipoles
of the solvent molecules, is the dominant
contribution to the total polarization. In this case Eq. (\ref{h1}) reduces to (see, c.f.,
Eqs. (\ref{GL3DU}) and (\ref{Fel}))

\be\label{h}
\begin{split}
 H&=\frac12 \sum_{\substack{{\bf r},{\bf r'}\\ {\bf r}\neq {\bf
r'}}}\sum_{i,j}n_i({\bf r})n_j({\bf r'})w\left(|{\bf r}-{\bf
r'}|\right)
\\&-\sum_{\bf r}\sum_iu_{\text
w}\delta_{z, a}n_i({\bf r})
\\&+\frac {1}{2} \int_ V\!d^3r^*\tilde\phi({\bf r^*})\tilde Q({\bf r^*})
\end{split}
\ee
where $\tilde Q({\bf r^*})=\frac{e}{a^3}\sum\limits_iq_i n^*_i({\bf
r^*})+\tilde\sigma\delta(z^*-a/2)$ is the local charge density where $n^*_i({\bf
r^*})=n_i({\bf r})$ for all ${\bf r^*}\in (a\mathbb{R})^3$ and ${\bf r}\in(a\mathbb{Z})^3$ with
$\max\left(|x^*-x|, |y^*-y|, |z^*-z|\right)\le a/2$; $\tilde\phi({\bf r^*})$ is the electrostatic
potential which can be obtained by solving the Poisson equation
\be\label{PoissonG}
-\varepsilon_0\tilde\nabla\cdot[\varepsilon(n^*_0({\bf r^*}))\tilde\nabla\tilde\phi({\bf
r^*})]=\tilde Q({\bf r^*},[n^*_\pm]),\ \
\ \ \
{\bf r^*}\in(a\mathbb{R})^3\cap V,
\ee
where $V$ is the volume of the fluid. For general permittivity profiles
$\varepsilon(n^*_0({\bf r^*}))$ no closed solution $\tilde\phi({\bf r^*})$ of Eq.
(\ref{PoissonG})
as a functional of $\varepsilon(n^*_0({\bf r^*}))$ and $\tilde Q({\bf r^*})$ is known, i.e., 
for each configuration $\{n_i({\bf r})\}$ the evaluation of
Eq. (\ref{h}) requires to solve the differential equation (\ref{PoissonG}) anew. It
has been proven, that models
including charges as in Eq. (\ref{h}) possess a proper thermodynamic limit for sequences of
finite-sized systems, which is
independent of the shape of the container,  provided that globally charge neutral
configurations $\{n_i(\bf r)\}$ are considered \cite{Lebowitz, Lieb}. Since the thermodynamic limit
is performed for  sequences of finite-sized systems the electrostatic potential $\tilde\phi({\bf
r^*})$ in Eq. (\ref{PoissonG}) vanishes at infinity ($|{\bf
r^*}|\to \infty$) \cite{Jackson}.

\subsection{Density functional}\label{DFT}

With a given expression for $\varepsilon (n_0({\bf r}))$ (see, c.f., Eq. (\ref{epsi})), Eq.
(\ref{h})
can be used directly for numerical analyses such as Monte Carlo simulations, provided an
efficient method to determine the electrostatic potential $\tilde\phi({\bf
r^*})$ for arbitrary permittivity profiles $\varepsilon (n_0({\bf r}))$ becomes available (see for
example Ref. \cite{Xu} for recent efforts in this direction). We leave this
challenging
task for future studies. Here, we consider a suitable mean field approximation which can be
formulated as to minimize a grand canonical density functional $\Omega[\{\rho_i({\bf
r})\}]$ \cite{Evans1979} of continuous and dimensionless occupation number distributions
$\rho_i({\bf
r})$ such that at the minimum  $\rho_i({\bf
r})=\rho_i^{eq}({\bf r})$ approximates the thermal average $\left\langle n_i(\bf r)\right\rangle$.

Application of the Bragg-Williams Approximation \cite{bellac, plischke, chaikin, safran} to the
model Hamiltonian in Eq. (\ref{h}) leads to the following grand canonical density functional:

\be\label{functional1}
\begin{split}
 \beta\Omega\left[\{\rho_i({\bf \bar r})\}\right]&=\sum_{{\bf \bar r}}\left[
\sum_i\rho_i({\bf \bar r})\ln{\rho_i({\bf \bar r})}\right.
\\&\left.+\Big(1-\sum_i\rho_i({\bf \bar
r})\Big)\ln{\Big(1-\sum_j\rho_j({\bf \bar r})\Big)}\right]\\
&+\frac12\beta\sum_{\substack{{\bf \bar r},{\bf \bar r'}\\ {\bf \bar r}\neq {\bf
\bar r'}}}\sum_{i,j}\rho_i({\bf \bar r})\rho_j({\bf \bar r'})w\left(|{\bf \bar r} -{\bf
\bar r'}|\right)
\\&-\beta\sum_{
{\bf \bar r}}\sum_iu_{\text
w}\delta_{\bar z, 1}\rho_i({\bf \bar r})-\beta\sum_{{\bf \bar r}}\sum_i\mu_i\rho_i({\bf \bar
r})
\\&+2\pi l_{B}\int_V\!d^3\bar
r^*\frac{\left({\bf
D}\left({\bf \bar r^*},[\rho^*_{\pm}]\right)\right)^2}{\varepsilon(\rho^*_0({\bf
\bar
r^*}))},
\end{split}
\ee
where $\beta=(k_BT)^{-1}$ is the inverse thermal energy and $\mu_i$ is the chemical potential
of species $i$, $\tilde l_{B}=l_B
a=e^2\beta/(4\pi\varepsilon_0)$ is the
Bjerrum length
in vacuum, ${\bf \bar r}={\bf r}/a$ are
 the dimensionless
 lattice positions,  ${\bf \bar r^*}={\bf r^*}/a$,  $\rho^*_i({\bf
\bar r^*})=\rho_i({\bf \bar r})$ for all ${\bf \bar r^*}\in \mathbb{R}^3$ and ${\bf \bar
r}\in\mathbb{Z}^3$ with
$\max\left(|\bar x^*-\bar x|, |\bar y^*-\bar y|, |\bar z^*-\bar z|\right)\le 1/2$. The
actual number densities of the components are given by
$\tilde \rho_i({\bf \bar r})=\rho_i({\bf \bar r})a^{-3}$. Charge neutrality demands
$\sum\limits_{\bf \bar
r}[\rho_+({\bf \bar r})-\rho_-({\bf \bar r})]=-\bar A\sigma$ where $A=MN=\bar Aa^2=\bar
M\bar Na^2$ is the substrate
area and $\sigma=\tilde\sigma/(ea^{-2})$; this constraint is implemented via a boundary condition
for ${\bf D}$ (see, c.f., Eq. (\ref{EN})). The first two terms of Eq. (\ref{functional1}) represent
the ideal gas or entropic
contribution $F_{id}$ to the Helmholtz free energy functional
$F\left[\{\rho_i({\bf \bar r})\}\right]=F_{id}\left[\{\rho_i({\bf \bar
r})\}\right]+F_{ex}\left[\{\rho_i({\bf \bar r})\}\right]$; the third and the fourth term
represent the non-electrostatic contribution to
$F_{ex}\left[\{\rho_i({\bf \bar r})\}\right]$,
which
follows from the first  and second term in Eq. (\ref{h}) and turns out to be equal to the
random phase approximation
(RPA) within density functional theory \cite{Evans1979}. This approximation is justified
because it has turned out that RPA
is reliable in the present situation of vanishing contrast between the non-electrostatic
interactions of the three species \cite{Bier2012}. The last term is the electrostatic
energy. Using SI units,  the
electrostatic field energy density, which enters into Eq. (\ref{functional1}), is given
by \cite{Jackson} 

\be
\frac12\tilde{\bf E} \cdot \tilde { \bf D}=-\frac 12 \frac{\tilde
{\bf D}^2}{\varepsilon_0\varepsilon}=\frac12
\frac{{\bf D}^2e^2}{\varepsilon_0\varepsilon a^4}=2\pi k_BT l_B \frac {{\bf D}^2}{\varepsilon
a^3}
\ee
and
\be
\frac12\tilde{\bf E} \cdot \tilde { \bf D}=-\frac12\tilde\nabla\tilde\phi\cdot\tilde{\bf
D}=\frac12\tilde\phi\left(\tilde\nabla\cdot\tilde{\bf D}\right)-\frac12\tilde
\nabla\cdot\left(\tilde \phi\tilde{\bf D}\right)\label{PE},
\ee
where  $\tilde {\bf E}=-\tilde \nabla\tilde \phi=\frac{\tilde {\bf D}}{\varepsilon_0\varepsilon}$ is
the actual
electric field, $\tilde \phi$ is the electrostatic potential and $\tilde
{\bf D}={\bf D}ea^{-2}$ is the actual electric
displacement generated by the ions and
the surface charge
density $\tilde\sigma=\sigma ea^{-2}$ , satisfying Gauss' law \cite{Jackson}

\be\label{GL3DU}
\tilde\nabla\cdot\tilde {\bf
D}=\tilde Q(\bf r^*),
\ee
so that ($\nabla=a\tilde\nabla$)
\be
\nabla \cdot {\bf D}\left({\bf\bar
r^*},[\rho^*_{\pm}]\right)=\sum_iq_{i}\rho^*_i({\bf \bar r^*})+\sigma\delta(\bar
z-1/2)\label{GL3D}.
\ee

Due to Eq. (\ref{PE}), the electrostatic contribution to the functional can be written as
\be\label{Fel}
F_{el}=\frac12\int d^3r^*\left[\tilde\phi({\bf
r^*})\left(\tilde\nabla\cdot\tilde{\bf
D}\right)-\tilde
\nabla\cdot\left(\tilde \phi\tilde{\bf
D}\right)\right],
\ee
where the last term leads to a vanishing surface contribution \cite{Jackson}, because the
thermodynamic limit is performed for sequences of finite-sized systems. Using Eq.
(\ref{GL3DU}) renders the last term in Eq. ({\ref{h}}).

Because the substrate potential depends only on $\bar z$, the minimum of
$\beta\Omega\left[\{\rho_i({\bf \bar r})\}\right]$ lies in the subspace of distributions
$\rho_i({\bf \bar r})$
which depend on $\bar z$ only. Therefore we write Eq. (\ref{functional1}) for the special
case
$\rho_i({\bf \bar r})=\rho_i(\bar z)$, i.e.,

\be\label{functional}
\begin{split}
 \frac{\beta\Omega\left[\{\rho_i(\bar z)\}\right]}{\bar A}&=\sum_{\bar z=1}^{\bar L}\left\{
\sum_i\rho_i(\bar z)\ln{\rho_i(\bar z)}\right.
\\&+\Big(1-\sum_i\rho_i(\bar z)\Big)\ln{\Big(1-\sum_j\rho_j(\bar
z)\Big)}\\
&-\beta u\sum_{ij}\big(\rho_{i}(\bar z)\rho_{j}(\bar z+1)+2\rho_i(\bar z)\rho_j(\bar
z)\big) \\
&\left.-\beta
u_{\text w}\sum_i\rho_i(\bar z)\delta_{\bar z,1}-\beta\sum_i\mu_i\rho_i(\bar z)\right\}
\\&+2\pi
l_{B}\int_{1/2}^{\bar L+1/2}\!d\bar z^*\frac{\left(D(\bar
z^*,[\rho^*_{\pm}])\right)^2}{\varepsilon(\rho^*_0(\bar
z))},
\end{split}
\ee
where $A=\bar
Aa^2$ is the substrate area so that $AL$ is the volume of the fluid ($L=\bar La$), and $\rho_i(\bar
L + 1)=0$. 

Gauss' law (Eq. (\ref{GL3D})) reduces to
\be\label{GL}
 \frac{dD(\bar z^*>1/2, [\rho^*_\pm])}{d\bar z^*}=\sum_iq_i\rho^*_i(\bar z^*)=\rho^*_+(\bar
z^*)-\rho^*_-(\bar
z^*),
\ee
where the last term in  Eq. (\ref{GL3D}) appears as a boundary condition to Eq. (\ref{GL}):
\be\label{BC}
D( \bar z^*=1/2, [\rho^*_\pm])=\sigma.
\ee
Since $\rho^*_{\pm}\in[0,1]$ are bounded, i.e., the densities $\rho^*_\pm$ do not exhibit 
$\delta$-like singularities, the
boundary
condition is determined entirely by the surface charge.

The density profiles $\rho_{\pm}(\bar z)$ have to fulfill global
charge neutrality, i.e.,
\be
\sum_{\bar z=1}^{\bar L}\left[\rho_+(\bar z)-\rho_-(\bar z)\right]+\sigma=0,
\ee
which according to the integrated Eq. (\ref{GL}) is equivalent to
\be\label{EN}
D\left(\bar z^*=\bar L+1/2,[\rho^*_{\pm}]\right)=0.
\ee

The relative permittivity $\varepsilon(\bar z^*)$ is taken to depend locally on the solvent
density $\rho^*_0(\bar z^*)$
through the
Clausius-Mossotti
expression \cite{Jackson}
\be\label{epsi}
\varepsilon(\rho^*_0(\bar z^*))=\frac{1+\frac{2\alpha}{3\varepsilon_0}\rho_0(\bar z^*)}{1-\frac{
\alpha}{
3\varepsilon_0 }
\rho^*_0(\bar z^*)},
\ee
where $\alpha$ is an effective polarizability of the solvent molecules. In the following its
value is
chosen such that $\varepsilon=60$ for $\rho_0=1$; this choice corresponds to a
mean value for liquid water
along the liquid-vapor coexistence curve.

As for a lattice model Eqs.
(\ref{h}) and (\ref{functional}) do not include the kinetic energy. The latter requires an
off-lattice description which leads to a density independent contribution to the
chemical potential of species $i$ so that
\be
\begin{aligned}
\mu_{i,phys}&= k_BT\ln(\tilde \rho_i \Lambda_i^3)+\mu_{ex},\\
&= k_BT\ln(\frac{\rho_i}{a^3} \Lambda_i^3)+\mu_{ex},\\
&=k_BT\ln(\rho_i)+\mu_{ex} +3k_BT\ln(\Lambda_i/a),\\
&=\mu_i+3k_BT\ln(\Lambda_i/a),
\end{aligned}
\ee
where $\Lambda_i=h/\sqrt{2\pi m_ik_BT}$ is
the thermal wavelength , $m_i$ is the particle mass, and
$\mu_{ex}$ is the excess chemical potential over the ideal gas contribution. This
gives rise to a density independent difference between the actual
physical
chemical potential $\mu_{i,phys}$ and the chemical potential $\mu_i$
of the lattice-gas model:  $\mu_{i,phys}-\mu_i=3k_BT\ln(\Lambda_i/a)$.

\subsection{Euler-Lagrange equations}
In order to obtain the equilibrium configuration, the density functional in Eq. (\ref{functional})
has to be minimized under the
constraints given by Eq. (\ref{BC}) and Eq. (\ref{EN}) \cite{Evans1979}. The variation of Eq. (\ref
{functional})
reads:
\begin{widetext}
\begin{equation}
\begin{aligned}
\label{variation}
 \frac{\beta\delta\Omega\left[\{\rho_i(\bar z)\}\right]}{\bar A}&=
\begin{aligned}[t]&\sum_{\bar
z=1}^{\bar L}\left\{
\sum_i\delta\rho_i(\bar z)\left[\ln{\rho_i(\bar
z)}-\beta\mu_i-\ln{\Big(1-\sum_j\rho_j(\bar z)\Big)}\right]\right. 
\\&-\beta u\sum_{ij}\!\Big(\delta\rho_{i}(\bar z)\rho_{j}(\bar z\!+\!1)+\rho_{i}(\bar
z)\delta\rho_{j}(\bar z\!+\!1)+2\delta\rho_i(\bar
z)\rho_j(\bar z)+2\rho_i(\bar
z)\delta\rho_j(\bar z)\Big)
\\&-2\pi l_{B}\int_{\bar z-1/2}^{\bar z+1/2}\!d\bar z^*\frac{\left(D(\bar
z^*,[\rho^*_{\pm}])\right)^2}{\left(\varepsilon\left(\rho^*_0(\bar
z^*)\right)\right)^2}\varepsilon'\left(\rho^*_0(\bar
z^*)\right)\sum_i\delta_{i,0}\delta\rho^*_i(\bar z^*)\\
&+\left.4\pi l_{B}\int_{\bar z-1/2}^{\bar z+1/2}\!d\bar z^*\frac{D(\bar
z^*,[\rho^*_{\pm}])}{\varepsilon(\rho^*_0(\bar
z^*))}\delta D(\bar z^*)-\beta
u_{\text w}\sum_i\delta_{\bar z,1}\delta\rho_i(\bar z)\right\} 
\end{aligned} \\
&=
\begin{aligned}[t]&\sum_{\bar
z=1}^{\bar L}\left\{
\sum_i\delta\rho_i(\bar z)\left[\ln{\rho_i(\bar
z)}-\beta\mu_i-\ln{\Big(1-\sum_j\rho_j(\bar z)\Big)}\right.\right. 
\\&\left.-\beta u\sum_{j}\Big(\rho_{j}(\bar z\!+\!1)+\sum_{\bar z'=1}^{\bar L}\rho_{j}(\bar
z')\delta_{\bar z,\bar z'+ 1}+2\rho_j(\bar z)+2\rho_j(\bar
z)\Big)-\beta
u_{\text w}\delta_{\bar z,1}\right]
\\&\left.-2\pi l_{B}\int_{\bar z-1/2}^{\bar z+1/2}\!d\bar
z^*\frac{\left(D(\bar
z^*,[\rho^*_{\pm}])\right)^2}{\left(\varepsilon(\rho^*_0(\bar
z^*))\right)^2}\varepsilon'(\rho^*_0(\bar
z^*))\delta_{i,0}\delta\rho_i^*(\bar z^*)-\int_{\bar z-1/2}^{\bar z+1/2}\!d\bar z^*\phi'(\bar
z^*)\delta D(\bar z^*)\right\} 
\end{aligned} 
\end{aligned}
\end{equation}
\end{widetext}

where $\phi(\bar z^*)=\beta e \tilde \phi(z^*)$ is the dimensionless
electrostatic potential which fulfills
\be\label{EDEP}
\begin{aligned}
\tilde D(z^*)&=\varepsilon_0\varepsilon \tilde E(
z^*)=-\varepsilon_0\varepsilon\frac{d\tilde
\phi(z^*)}{dz^*},\\
ea^{-2}D(\bar z^*)&=-\varepsilon_0\varepsilon \frac{1}{a}\frac{d}{d\bar z^*}\left(\frac{\phi(\bar
z^*)}{\beta e}\right),\\
 D(\bar z^*)&=-\frac{\varepsilon}{4\pi l_{B}}\phi'(\bar z^*).
\end{aligned}
\ee

Upon integrating by parts the last term in
Eq.
(\ref{variation}), by using Eq. (\ref{BC}) so that $\delta D(\bar
z^*=1/2)=0$ and Eq. (\ref{EN}) so
that $\delta D(\bar
z^*=\bar L+1/2)=0$,
with $\delta D'(\bar z^*)=\sum_iq_i\delta\rho^*_i(\bar z^*)$ due to Eq. (\ref{GL}), and
$\delta\rho^*(\bar
z^*)=\delta \rho_i(\bar z)$ for all $\bar z^*\in \mathbb{R}$ and $\bar z\in \mathbb{Z}$ with
$\max(|\bar z^*-\bar z|\le1/2)$ we obtain
the
following three coupled
Euler-Lagrange equations for $\bar z\in\{1,\dots,\bar L\}$
\be\label{ELE}
\begin{split}
 & \ln{\rho_i(\bar z)}-\mu^*_i-\beta u_{\text w}\delta_{\bar z,1}
-\ln{\Big(1-\sum_j\rho_j(\bar
z)\Big)}\\
&-\frac{1}{3T^*}\sum_j\left(4\rho_{j}(\bar z)+\rho_j(\bar z+1)+\rho_{j}(\bar
z-1)\right)\\
&+q_i\int_{\bar z-1/2}^{\bar z+1/2}\!d\bar z^*\phi(\bar z^*)\\
&-2\pi
l_{B}\int_{\bar z-1/2}^{\bar z+1/2}\!d\bar z^*\frac{\left(D(\bar
z^*,[\rho^*_{\pm}])\right)^2}{\left(\varepsilon(\rho^*_0(\bar
z^*))\right)^2}\varepsilon'\left(\rho^*_0(\bar
z^*)\right)\delta_{i,0}=0
\end{split}
\ee
with $i,j=0,+,-$, where $q_ie$ is the electric charge of component $i$ and
$T^*=\frac{1}{3\beta u}$ is the reduced temperature and $\mu^*_i=\beta \mu_i$. At the wall the
convention $\rho_j(0)=0$ is
used. The integrals in Eq. (\ref{ELE}) are approximated by
\begin{multline}
\int_{\bar z-1/2}^{\bar z+1/2}\!d\bar z^*f(\bar z^*)\approx\\
\left(\left(\bar
z+\sfrac12\right)-\left(\bar z-\sfrac12\right)\right)f\left(\frac{\left(\bar
z+\sfrac12\right)+\left(\bar
z-\sfrac12\right)}{2}\right)=\\=f(\bar z).
\end{multline}
 
For given chemical potentials
$\mu_i$ these coupled equations can be solved numerically by an
iterative algorithm. The values of the chemical potentials $\mu_i$ considered
here correspond to those for the bulk gas phase of the system. For each iteration the electrostatic
potential $\phi(\bar z^*)$ must be calculated by solving Poisson's equation (see Eqs. (\ref{GL}) and
(\ref{EDEP}))
\be
\frac{d}{d\bar z^*}(\varepsilon(\rho^*_0(\bar z^*))\phi'(\bar z^*))=-4\pi l_B\sum_iq_i\rho^*_i(\bar
z^*),
\ee 
ensuring global charge neutrality at each step.

\subsection{Wetting films}

The wetting behavior can be transparently inferred from the constrained surface contribution
$\Omega_s\left(l\right):=
(\Omega[\{\rho_i^{(l)}\}]-\Omega_b)/A$ to the grand potential \cite{Dietrich1988}, where $\Omega_b$
is the bulk
contribution to the grand potential and the density profiles $\rho_i^{(l)}$ are
the solutions
of
the Euler-Lagrange equations (\ref{ELE}) for a prescribed film thickness $\tilde l=la$ defined
as
\be\label{ft}
l=\frac{\Gamma}{\rho_{0,\mathrm{l}}-\rho_{0,g}},
\ee 
where $\tilde \Gamma=\int_0^\infty{dz\left(\tilde \rho_0(z)-\tilde \rho_0(\infty)\right)}=\Gamma
a^{-2}$ is
the excess adsorption (or coverage) of the substrate
by the solvent
and $\rho_{0,\mathrm{l}}$ and $\rho_{0,g}$ are the corresponding bulk number densities of the liquid
and
the gas phase,
respectively. In
order to obtain
$\Omega_s\left(l\right)$ by using a Lagrange multiplier we have minimized $\Omega\left[
{\rho_i(z)}\right]$ under the
constraint
\be
\sum_{\bar z=1}^{\infty}(\rho_0(\bar
z)-\rho_{0,b})=\Gamma=l(\rho_{0,\mathrm{l}}-\rho_{0,g}),
\ee
where $\rho_{0,b}$ is the number density of the bulk gas phase in units of $a^{-3}$.
\subsection{Choice of parameters}
If one chooses the lattice constant $a$ to be equal to
$4$\AA , the maximal density $1/a^3$ lies between the densities
for liquid
water at the triple point and at the critical point. Accordingly, the choice $l_{B}=400$ 
corresponds
to
$T\approx417$ K. This temperature lies between the triple point temperature of 273 K and the
critical
point temperature of 647 K
for water. In our units $1~\text{mM}=10^{-3}~\text{mol/L}$ corresponds to $\rho_i=\tilde \rho_ia^3
=3.9\times10^{-5}$.

For our calculation we have used values for the reduced surface charge density $\sigma$ in
the
range between 0
and
$10^{-2}$. For $a=4$\AA~ the latter value corresponds to 1 $\mu$C/cm$^2$. Such values are within
the range of measured surface charge densities of silicon nitride at two different concentrations of
the background electrolyte NaCl (1~mM,
10~mM) determined by
potentiometric pH
titration \cite{Sonnefeld199627},  which is a common method to
determine the unknown concentration of an identified substance and to estimate the surface
charge of a solid by comparing the titration of the solution with solid against the titration of the
same
solution without solid.

These consideration indicate that the values of the reduced
substrate surface charge densities $\sigma$ and ionic strengths $I$ considered in the
following are within the range of values for which Poisson-Boltzmann theory, i.e., mean-field
theory for the
electrostatic interaction, shows quantitative agreement with corresponding Monte Carlo simulations
\cite{Torrie1979}.
The former is essentially identical to the theory used to describe the ions in
Eq. (\ref{functional}) if
one neglects the effect of nonzero ion size, which is weak for the considered dilute electrolyte
solutions.

\section{Results and Discussion}\label{res}
\subsection{Bulk Phase Diagram}\label{bulk}

In the bulk, the number densities $\rho_i$ of the fluid are spatially constant and from the
requirement of
local charge neutrality it follows that $\rho_+=\rho_-=I$, where $I$ is the so-called
ionic strength for monovalent ions.
Under these conditions the density functional given by Eq. (\ref{functional})
reduces to

\begin{multline}\label{bulk1}
\frac{\beta \Omega[\{\rho_i\}]}{\bar V}=\rho_0(\ln{\rho_0}-\mu^*_0)
+I(2\ln{I}-\mu^*_I)\\
+(1-\rho_0-2I)\ln{(1-\rho_0-2I)}-\frac{1}{T^*}(\rho_0+2I)^2,
\end{multline}
where $\mu^*_I=\mu^*_++\mu^*_-$ and $\bar V=V/a^3$ ($V$ is the volume of the fluid). The last
term in Eq. (\ref{functional}) vanishes because in the bulk $D=0$ due to Eq. (\ref{GL}). The
Euler
Lagrange equations (\ref{ELE}) read
\be
\begin{aligned}
 \ln{\rho_0}-\mu^*_0 -\ln{(1-\rho_0-2I)}-\frac{2}{T^*}(\rho_0+2I)&=0\\
2\ln{I}-\mu^*_I -2\ln{(1-\rho_0-2I)}-\frac{4}{T^*}(\rho_0+2I)&=0.
\end{aligned}
\ee
\begin{figure}[!t]
 \includegraphics[width=8cm,clip=true]{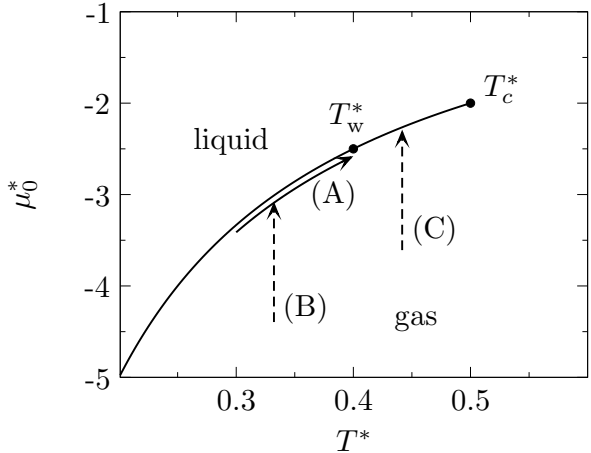} 
 \caption{ Bulk phase diagram $\mu_{0,co}(T)$ of liquid-gas \textit {co}existence according to Eq.
(\ref{bulk1}) in the
$\mu^*_0-T^*$ plane for the salt-free ($I=0$)
case of a pure solvent. If the wetting transition temperature
$T^*_{\text w}$ is above the triple point $T^*_t\simeq0.21$ (for water), three types of paths
(A), (B), and (C) are used
to
study the wetting behavior of our model. (A)
is a path along
gas-liquid coexistence on the gas side whereas along the paths (B) and (C) two-phase coexistence is
approached along
isotherms leading to incomplete (B) and complete (C) wetting, respectively.}
\label{wet}
\end{figure}


For a given ionic strength $I=\rho_{\pm}^{(\mathrm l)}$ in the liquid phase of the solution, the
liquid-gas
coexistence curves, i.e.,
the solvent density in the liquid phase of the solution and the coexisting densities of the ions and
of the solvent in the gas phase of the solution, are determined by the equality of the chemical
potentials $\mu_0$ and $\mu_I$ and of the pressure $p$: 
\be
\begin{aligned}\label{coex}
\mu_0[\{\rho_i^{(g)}\},T^*] & =\mu_0[\{\rho_i^{(\mathrm l)}\},T^*],\\
\mu_I[\{\rho_i^{(g)}\},T^*] & =\mu_I[\{\rho_i^{(\mathrm l)}\},T^*],\\
p[\{\rho_i^{(g)}\},T^*] & =p[\{\rho_i^{(\mathrm l)}\},T^*].\\
\end{aligned}
\ee

For $I=0$ the resulting phase diagram can be determined analytically and is plotted in
Fig.~\ref{wet}.
The reduced critical
temperature is $T^*_c(I=0)=0.5$ and the critical number density is $\rho_{0,c}(I=0)=0.5$. For
$I\neq0$ the
binodal
curves are                                  
determined numerically and the critical points are obtained by determining the
maximum of the
corresponding spinodal curves. Within the present model the reduced critical
temperature $T^*_c$ is
independent of $I$ whereas
$\rho_{0,c}(I)=0.5-2I$. In agreement with experimental evidence \cite{Seah1993} the shift of the
binodal curves is negligibly small for ionic strengths
up to 10 mM, i.e., $I\le3.9\times10^{-4}$.

\subsection{Wetting}\label{wetting}
\subsubsection{Salt-free solvent}\label{sal-free}

We first consider the case $I=0$, in which our model reduces to the lattice-gas
model studied by
Pandit et al. \cite{Pandit1982, Pandit1983}. In that case, the Euler-Lagrange
equations in Eq. (\ref{ELE}) reduce to
\begin{multline}
\ln{\rho_0(\bar z)}-\ln{[1-\rho_0(\bar z)]}-\mu^*_0-\beta u_{\text
w}\delta_{1,\bar
z}\\-\frac{1}{3T^*}
\left[4\rho(\bar z)+\rho(\bar z+1)+\rho(\bar z-1)\right]=0,
\end{multline}
and the ratio $u_{\text w}/u=3T^*\beta u_w$
controls the wetting and drying transitions. For $u_{\text w}/u>1$ the
substrate is so strong that it is already wet at
$T^*=0$; in the range $0.5<u_{\text w}/u<1$ there is a wetting transition at
$T^*_{\text w}>0$; and in
the
parameter
range $0\leq u_{\text w}/u<0.5$ a drying transition occurs. Depending on the value of the
ratio $u_{\text
w}/u$ one
observes
layering transitions, i.e., one can distinguish the number of discrete layers which are forming
upon reaching thick films. The
transition
from $n$ to $n+1$ layers is first order and shows up as a jump in the film
thickness $l$. The loci of these
discontinuities  are layering transition lines, each ending at a critical point
$T^*_{c,n}$. For large $n$, $T^*_{c,n}$ approaches the roughening transition. However, within the
present mean-field theory $T^*_{c,n}$ approaches $T^*_c$. Since
layering transitions should only occur along  or near the melting
curve or the sublimation line, these layering transitions are a
special feature of the lattice-gas model used to describe the liquid and gas
phases \cite{Dietrich1988}.

\begin{figure}[!t]
  \includegraphics[width=7cm]{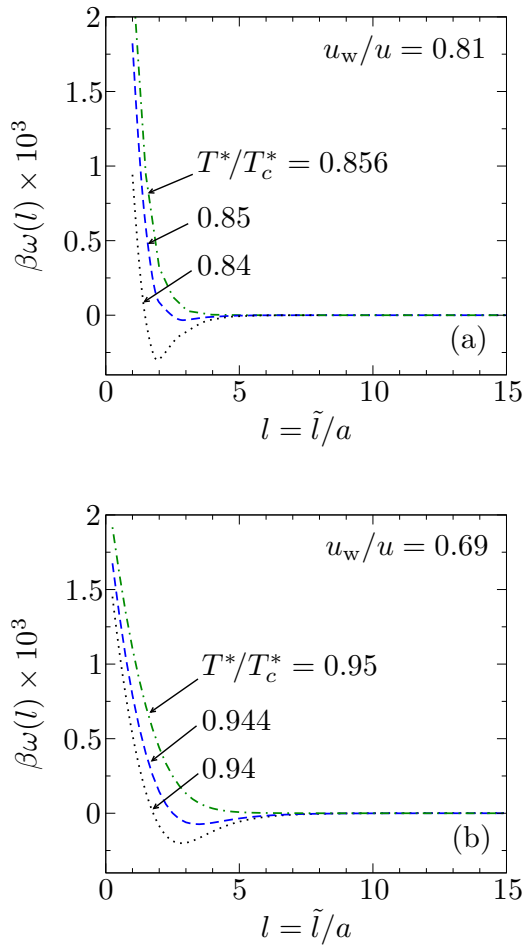}
   \caption{ Effective interface potential
$\omega(l)=\Omega_s(l)-\gamma_{g,l}-\gamma_{l,s}$ at two-phase coexistence as a
function of the thickness $\tilde l=la$ of
the adsorbed liquid
film for three temperatures in the salt-free
case ($I=0$) for $u_{\text
w}/u=0.81$ (a) and $u_{\text w}/u=0.69$ (b) . In both cases  $\omega(l)$ exhibits
only a single
minimum, the position of which
diverges
continuously as $ T^*\to T^*_{\text w}$. Accordingly, the system undergoes
critical wetting at
$T^*_{\text w}\simeq0.856T^*_c$ for $u_{\text w}/u=0.81$
and at
$T^*_{\text w}\simeq0.95T^*_c$ for $u_{\text w}/u=0.69$.}
\label{transition}
\end{figure}

We have carried out calculations in the parameter range $0.5<u_{\text w}/u<1$. A wider range
of the parameter $u_{\text
w}/u$  was studied thoroughly by
Pandit et al. \cite{Pandit1982, Pandit1983}. Figure
\ref{transition}
shows the effective interface potential
$\omega(l)=\Omega_s(l)-\gamma_{g,l}-\gamma_{l,s}$ for
three different temperatures along a path at
coexistence [path (A) in Fig.~\ref{wet}] for the rather arbitrarily chosen values $u_{\text
w}/u=0.81$ and $u_{\text
w}/u=0.69$. Here
$\gamma_{g,l}$ and $\gamma_{l,s}$ are the gas-liquid and liquid-substrate
interfacial tensions,
respectively, such that by construction at two-phase coexistence $\omega(l\to \infty)=0$. The
equilibrium thickness of the liquid film is given by
the position of the global minimum of $\omega(l)$. If $l=\infty$ is the global
minimum of $\Omega_s(l)$ the system is wet. In this case, the gas-substrate
surface tension is given by
$\gamma_{g,s}=\Omega_s(l=\infty)=\gamma_{g,l}+\gamma_{l,s}$.

In the two cases which we have considered in Fig.
\ref{transition}, $\omega(l)$ exhibits only a single
minimum, the position of which
diverges
continuously or via steps of finite size as $ T^*\to T^*_{\text w}$. For $T^*>T^*_{\text w}$ the
position of
the minimum is
$l=\infty$ and the system is wet. The wetting
transition is second order and occurs at the temperature $T^*_{\text w}\simeq0.856T^*_c$ for
$u_{\text w}/u=0.81$
and at
$T^*_{\text w}\simeq0.95T^*_c$ for $u_{\text w}/u=0.69$. Within the present model, in which all
interactions are of the nearest-neighbor type only for the pure solvent, the system exhibits a
second-order wetting transition in the entire parameter range $0.5<u_{\text w}/u<1$. This
observation is compatible with corresponding Monte Carlo simulations of the Ising model on a cubic
lattice 
\cite{Binder1988, Binder1989}. However, the order of wetting transitions depends sensitively on the
range of interactions as well as on whether a continuous or a lattice model is considered. For a
continuous analogue of the present model, Pandit et al. \cite{Pandit1983} found
a second-order
wetting transition only for $0.5<u_{\text w}/u\lesssim0.7$ but a first-order one for $u_{\text
w}/u\gtrsim0.7$.  

Moreover, lattice-gas models with short-ranged particle-particle interactions and long-ranged
substrate
potentials were studied by de Oliveira and Griffiths \cite{Oliveira1978} and Ebner \cite{Ebner1980,
Ebner1981}. In Ref.  \cite{Oliveira1978} complete wetting in a system with $T_\text w=0$ was
studied within mean field theory. Ebner reported $T_\text w=0$ or
a first-order wetting transition depending on the strength of the substrate potential
\cite{Ebner1980}
and studied the same interaction potentials as the ones used in Refs. \cite{Oliveira1978, Ebner1980}
applying Monte Carlo
simulations \cite{Ebner1981}. Finally, in systems in which both the particle-particle
interactions and the substrate
potentials are long-ranged, critical (i.e., second-order) and first-order wetting can occur for 
suitable choices of the interaction
potentials \cite{Dietrich1985, Ebner1985}.

\begin{figure}[!t]
\includegraphics[width=7cm]{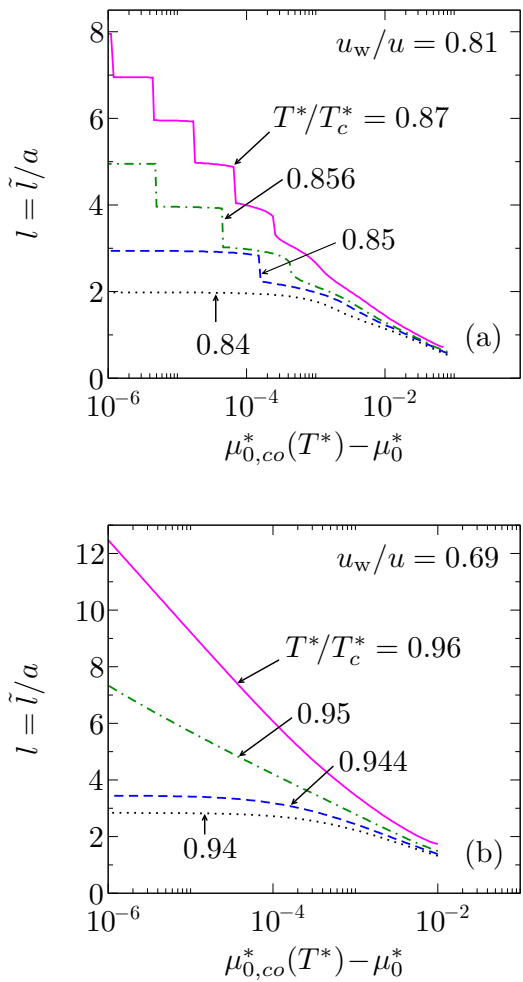} 
\caption{Film thickness $l=\tilde l/a$ in units of the lattice constant $a$ as a function of
undersaturation
$\mu^*_{0,co}(T^*)\!-\!\mu^*_0$ for
the salt-free case ($I=0$). Gas-liquid coexistence $\mu^*_{0,co}(T^*)$ is approached from the gas
phase.
(a)
$u_{\text
w}/u=0.81$: for
$T^*<T^*_{\text w}=0.856 T^*_c$ the system is partially wet and, if at all, there is a finite number
of layering
transitions;
for $T^*>T^*_{\text w}$ the isotherms exhibit an unlimited number of layering
transitions as
$\mu^*_{0,co}(T^*)\!-\!\mu^*_0\to0$ and the first few layering transitions are rounded
because for this temperature $T^*>T^*_{c,n}$. (b)
$u_{\text w}/u=0.69$: the film thickness diverges logarithmically for
$T^*>T^*_{\text w}=0.95T^*_c$, while
it reaches a finite value at coexistence for $T^*<T^*_{\text w}$. In (b) there are no layering
transitions. Note that with $T^*_c(I)=\frac12$ one has
$\mu^*_{0,co}(T^*)\!-\!\mu^*_0=\frac{2/3}{T^*/T^*_c}\left[\frac{\mu_{0,co}
(T^*)-\mu_0}{u}\right]$.} 
\label{logsolvent}
\end{figure}

The film thickness $l=\tilde l/a$ as function of $\mu^*_{0,co}(T^*)-\mu^*_0$, when bulk
coexistence
$\mu_{0,co}(T^*)$ (see Fig. \ref{wet}) is
approached along four 
isotherms from the gas phase [paths of type (B) and (C) in Fig.~\ref{wet}], is plotted
in Fig.
\ref{logsolvent}. In the case $u_{\text w}/u=0.81$ (Fig. 3(a)) the isotherms exhibit
vertical steps at the aforementioned layering
transitions. Above $T^*_{\text w}$, i.e., when the substrate is completely wet
at coexistence, the
isotherms exhibit an unlimited number of such steps as $\mu^*_{0,co}(T^*)-\mu^*_0$
approaches zero, while for 
$T^*<T^*_{\text w}$ there is only a finite number of steps. For $u_{\text
w}/u=0.69$ (Fig. 3(b)) layering
transitions do not occur and the film thickness diverges logarithmically for
$T^*>T^*_{\text w}$,
while for $T^*<T^*_{\text w}$ it reaches a finite value at coexistence.

\subsubsection{Electrolyte solution}\label{electrolyte}
Within the above concepts we now focus on the influence of the ionic strength $\tilde I=Ia^{-3}$
and of the surface charge
density $\tilde \sigma=\sigma e a^{-2}$ on the wetting behavior of systems with $u_{\text
w}/u=0.81$ or $u_{\text
w}/u=0.69$. If the substrate is neutral ($\sigma=0$), the
addition
of salt changes neither the order
nor the transition temperature of the wetting
transition, i.e., there is a second-order wetting transition at the wetting temperature
$T^*_\text{w}$
as discussed in the previous Subsubsec. \ref{sal-free}. This is expected because within our model
all
particles have the same
size, the ions have
the same absolute charge, and the strength of the particle-particle and of the
substrate-particle
nearest-neighbor
interactions are
the same for all three species. Hence local charge neutrality ($\rho_+(\bar
z)=\rho_-(\bar z)$) holds due to the exchange symmetry with respect to the ionic components. This
implies that there is no
electric field ($D(\bar z)=0$). If the surface charge becomes non-zero, the
order of the wetting transition changes from second order ($\sigma=0$) to
first order ($\sigma\neq0$) for all
values of the charge density $\sigma$ and
ionic strength $I$ studied here, with
$\sigma =2\times10^{-5}\ (\text{i.e.,\ }\tilde\sigma\approx0.002~\mu\text{C}/\text{cm}^2)$ as the
smallest non-zero value considered. This result is in agreement with previous
studies. The influence of ionic
solutes on the order of the wetting
transition was studied in Ref.
\cite{Denesyuk2004} by using Cahn's phenomenological theory and
in Ref. \cite{Oleksy2009} by using density functional theory for an explicit solvent
model for an ionic solution. Both studies suggest that electrostatic interactions favor first-order
wetting.

\begin{figure}[!t]
  \includegraphics[width=7cm]{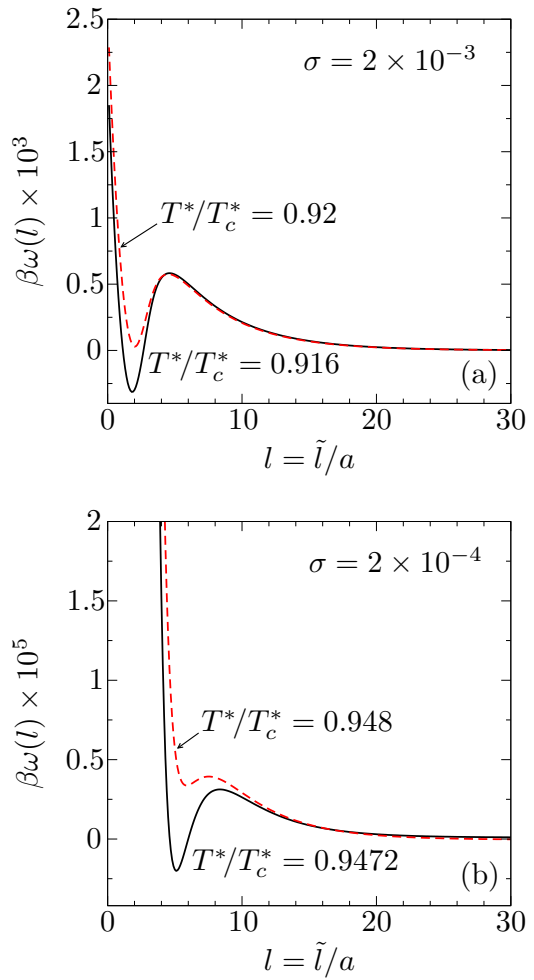}
   \caption{Effective interface potential
$\omega(l)$ at gas-liquid coexistence as
function of the thickness $l=\tilde l/a$ of
the liquid
film for $u_{\text w}/u=0.69$, $I=3.9\times10^{-5}$ ($\tilde I=1$mM), and
$\sigma=2\times10^{-3}$ ($\tilde\sigma=0.2\mu \text{C/cm}^2$) in (a)
and $\sigma =2\times10^{-4}$
($\tilde\sigma=0.02\mu \text{C/cm}^2$) in (b) for two temperatures in each case.
The effective interface
potential $\omega(l)$ has two local minima (one at $l<\infty$ and one at $l=\infty$)
which 
have the
same depth at $T^*_{\text w}$. Accordingly, for both surface charge densities $\sigma$ the
system undergoes a first-order
wetting transition.}
\label{first}
\end{figure}

Figure~\ref{first} shows examples of the effective interface potential $\omega(l)$
in the case of
non-zero surface charge densities, $\sigma = 2\times10^{-3}$  and
$\sigma = 2\times10^{-4}$,
 for two temperatures and at bulk coexistence [see path (A) in Fig.
\ref{wet}]. In both cases, $\omega (l)$ has two local minima. For
$T^*<T^*_{\text w}$ the global minimum corresponds to a thin film whereas
for $T^*>T^*_{\text
w}$ the film is macroscopically thick.
At the
wetting transition 
temperature $T^*_{\text w}$ the two minima correspond to the same value of the effective
interface potential
$\omega(l)$. Accordingly, at \Tws~the film thickness jumps discontinuously from a finite value
below $T^*_{\text w}$ to a macroscopic one above \Tws~ so that the system
undergoes a first-order wetting
transition. If $\sigma$ is decreased the height of the barrier in $\omega(l)$ at the wetting
temperature
$T^*_\text{w}$ decreases and the minimum close to the wall is shifted to larger thicknesses (Fig.
\ref{first}(b)). In the case
$\sigma=0$, $\omega(l)$ has only a single minimum, like in the salt-free 
case (see Fig.~ \ref{transition}), corresponding to a second-order wetting transition.

\begin{figure}[!thb]
   \includegraphics[width=8cm]{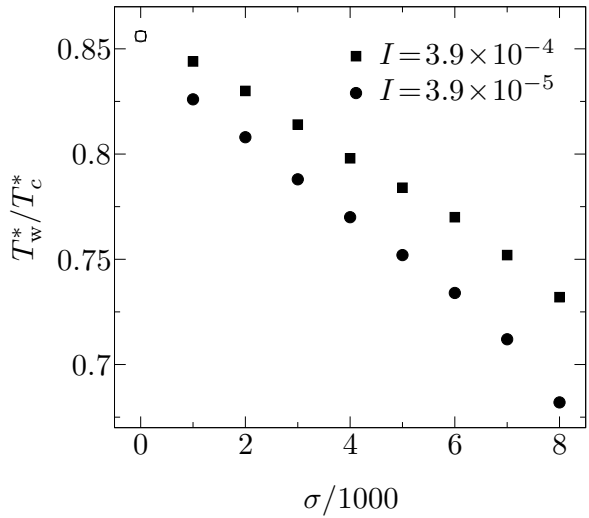}  
   \caption{Wetting transition temperature $T^*_{\text w}$ as a function of the substrate surface
charge 
density $\sigma=\tilde \sigma a^2/e$ for $u_{\text
w}/u=0.81$.
The two types of symbols correspond to distinct values of the ionic
strength $I=\tilde Ia^{3}$ in the bulk liquid phase ($\bullet$ for $I=3.9\times10^{-5}$ ($\tilde
I=1$mM) and $\blacksquare$ for $I=3.9\times10^{-4}$ ($\tilde
I=10$mM)). Filled symbols
correspond to first-order wetting transitions, while the empty one at $\sigma=0$ corresponds to a
second-order wetting transition, with the corresponding wetting transition temperature being
independent of $I$.}
\label{Tw-sig}
\end{figure}

In Fig.~\ref{Tw-sig} the wetting transition temperature
is plotted as function of the
surface charge density for two values of the ionic strength and for $u_{\text
w}/u=0.81$. As $\sigma=\tilde \sigma a^2/e$
is
increased, the wetting transition
temperature $T^*_{\text w}$ decreases due to the strengthening of the
substrate-fluid attraction as the substrate is charged up. For $\sigma\neq0$ the
system with a smaller 
ionic strength $I$ has always the lower wetting transition temperature $T^*_{\text w}$ because
in this case the 
screening of the electrostatic forces of the substrate
is reduced making them effectively stronger which favors wetting. As already
mentioned above, within our model for $\sigma=0$
the wetting transition
temperature
is independent of the ionic strength $\tilde I=Ia^{-3}$. The trend is the same for
$u_{\text w}/u=0.69$. In the case of first-order
wetting transitions these results are in qualitative
agreement with Ref. \cite{Oleksy2009}. However, the off-lattice model used therein exhibits also
second-order wetting transitions (see the discussion above in Subsubsec. \ref{sal-free}), for which
$T^*_{\text w}$ is a non-monotonic function of $\sigma$. 
 
\begin{figure}[!t]
   \includegraphics[width=8cm]{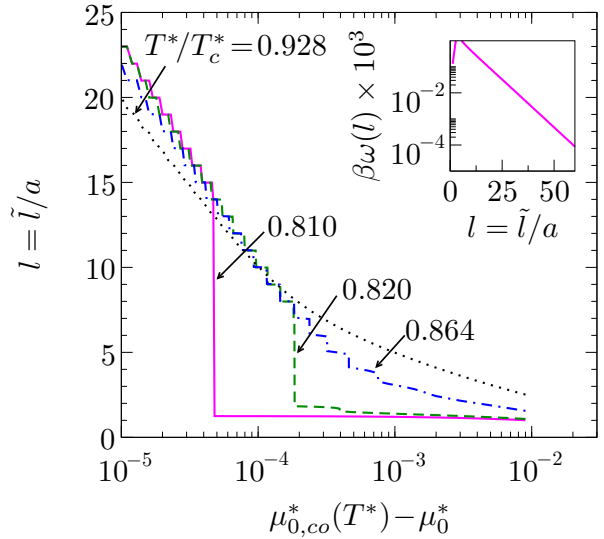}  
   \caption{The film thickness $l=\tilde l a$ (Eq. (\ref{ft})) as a function of undersaturation
$\mu^*_{0,co}(T^*)\!-\!\mu^*_0$ along four
different isotherms for $u_{\text w}/u=0.81$, $I=3.9\times10^{-5}$ ($\tilde I=1$mM), and $\sigma
=2\times10^{-3}$ ($\tilde \sigma=0.2 \mu\text{C}/\text{cm}^2$)
exhibits
a large but finite jump (corresponding to more
than one monolayer) when the prewetting line is crossed and small jumps
when the various layering
transition lines are crossed. The film thickness increases for small undersaturation as
$l\sim\ln(\mu^*_{0,co}(T^*)\!-\!\mu^*_0)$  where
$\mu^*_{0,co}(T^*)\!-\!\mu^*_0=\tfrac{2/3}{T^*/T^*_c}\left[\tfrac{\mu_{0,co}
(T^*)-\mu_0}{u}\right]$. The inset displays the  corresponding asymptotic behavior of the
effective interface potential $\omega(l)\sim \exp(-2\kappa l)$ where
$\kappa=\sqrt{8\pi l_BI/(\varepsilon(\rho_0^{(\mathrm{l})})}$ is the inverse Debye length.} 
\label{layersandpw}
\end{figure}

\begin{figure}[!t]
   \includegraphics[width=7cm]{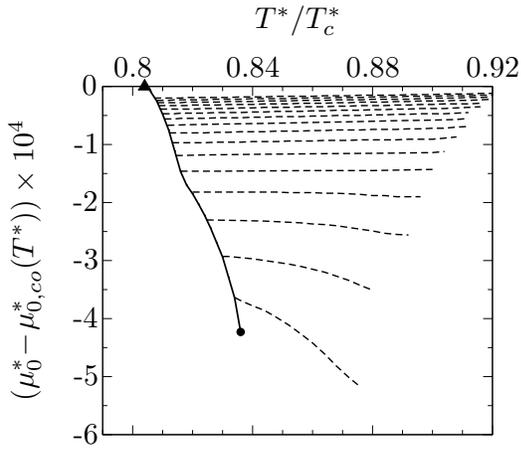}  
   \caption{Surface phase diagram for $u_{\text w}/u=0.81$ and
$\sigma=2\times10^{-3}$ ($\tilde \sigma=0.2 \mu\text{C}/\text{cm}^2$)
. The full line is the prewetting line attached to $T_\text{w}^*=0.864\,T^*_c$
($\blacktriangle$) and ending at the prewetting critical point ($\bullet$). The
dashed
lines
correspond to
layering transition lines. They end at layering critical points $T^*_{c,n}$ (located at the end of
the dashed lines without being indicated separately), which within the present mean-field theory
accumulate for $n\to \infty$ at $T^*_c$ instead of at the roughening transition temperature of the
gas-liquid interface on  the lattice.}
\label{phase-diagram}
\end{figure}

Since the wetting transitions for $\sigma\neq0$ are first order, there is
a prewetting line
associated with them. The prewetting line is attached tangentially to the gas-liquid
coexistence line at
the wetting temperature $T^*_{\text w}$ and bends away from coexistence, marking the loci of a
finite
discontinuity in film
thickness $l=\tilde l/a$. The discontinuity upon crossing the prewetting line gets
smaller as one moves further away from coexistence
and it vanishes at the prewetting critical point. Figure~\ref{layersandpw} shows
the film thickness $l=\tilde l/a$ for four different isotherms as a function of
undersaturation $\mu^*_{0,co}(T^*)-\mu^*_0$ for
$u_{\text w}/u=0.81$ and $\sigma=2\times10^{-3}$ ($\tilde \sigma=0.2 \mu\text{C}/\text{cm}^2$). The
film thickness increases for small undersaturation as $l\sim\ln(\mu^*_{0,co}(T^*)-\mu^*_0)$.
Accordingly, $\omega(l)\sim \exp(-2\kappa l)$, where
$\kappa=\sqrt{8\pi l_BI/\varepsilon(\rho_0^{\mathrm{l}})}$ is the inverse Debye length (see inset
of Fig. \ref{layersandpw}).
This is in
agreement with Refs. \cite{Kayser1988} and \cite{Denesyuk2004} for wetting of solvents with added
salt. In contrast, for wetting films of solvents without addition of salt, i.e., with counterions
only, one has $l\sim(\mu_{co}-\mu)^{-1/2}$ and $\omega(l)\sim
l^{-1}$ \cite{Langmuir1938,Kayser1986, Denesyuk2004, Derjaguin1974}. In order to obtain this result,
Eqs. (\ref{h}) and (\ref{functional}) have to be modified to consider only solvent particles and
counterions but leaving out coions. In
addition to the
finite thin-thick jumps in film thickness $l$ when crossing the prewetting line we observe
first-order layering
transitions similar to those found in the
salt-free case for $u_{\text w}/u=0.81$ (see Fig.~\ref{logsolvent}). The addition of
the electrostatic interaction leads to a
series of triple points where the layering transition lines meet the prewetting line, as shown in
the surface phase diagram in Fig.
\ref{phase-diagram}. A similar phase diagram was found by Ebner \cite{Ebner1980}
using a lattice-gas
model for a one-component fluid in which the fluid particles interact among each other via a
Lennard-Jones (6-12)
potential and a fluid
particle interacts with the substrate via a (9-3) potential. This is also in line with the
prediction by Pandit et al.
\cite{Pandit1982} for a substrate of intermediate strength, i.e., for $0.5<u_{\text
w}/u<1$, with interactions ranging beyond nearest
neighbors.

\begin{figure}[!t]
   \includegraphics[width=8cm]{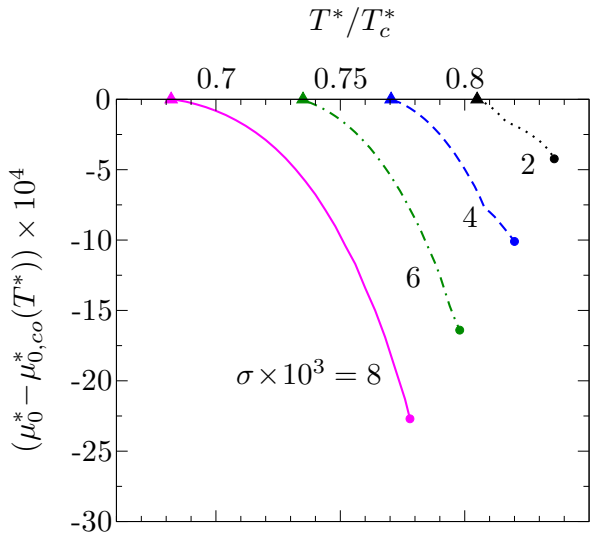}  
   \caption{Prewetting lines for four values of the surface charge density $\sigma=\tilde \sigma
a^2/e$ with ionic
strength $I=3.9\times10^{-5}$ ($\tilde I=1$mM) in the bulk liquid phase and
for $u_{\text
w}/u=0.81$. The locations of the wetting transitions ($\blacktriangle$) and of the prewetting
critical points ($\bullet$) are given in Table
\ref{tab:1}.}
\label{prewet}
\end{figure}

In the case  $u_{\text w}/u=0.81$ and
for fixed ionic strength $I$ we have studied the prewetting lines for various
values of the surface charge density $\sigma$.  Figure~\ref{prewet}
shows the prewetting
lines for
ionic strength $I=3.9\times10^{-5}$ ($\tilde I=1$mM) and for four values of $\sigma$. One can see
clearly that as
$\sigma$ 
decreases, the wetting  temperature $T^*_{\text w}$ rises and the prewetting
line becomes shorter. This is
in agreement with the fact that in the limit $\sigma \to 0$ the wetting transition turns second
order.
The values of the prewetting critical points for the lines shown in Fig.~\ref{prewet} are given in
Table \ref{tab:1}. 

\begin{table}[!t]
   \begin{tabular}{cccc}
\hline
\hline
\ $\sigma=\tilde \sigma a^2/e$ \ \ &\ \ $T^*_\text{w}/T^*_c$\ \ &\
\ \ $T^*_{p\text{w},c}/T^*_c$\ \ &\
\ $\mu^*_{0,co}\left(T^*_{p\text{w},c}\right)-\mu^*_{0,p\text{w},c}$\ \\ \hline
$2\times10^{-3}$& 0.804 & 0.836 & $4.23\times10^{-4}$ \\
$4\times10^{-3}$& 0.77 & 0.82 & $1.01\times10^{-3}$ \\
$6\times10^{-3}$& 0.734 & 0.798 & $1.60\times10^{-3}$ \\
$8\times10^{-3}$& 0.682 & 0.778 & $2.27\times10^{-3}$ \\
\hline\hline
  \end{tabular}
\caption{Prewetting critical points $(T^*_{p\text{w},c},\mu^*_{0,p\text{w},c})$ for the prewetting
lines shown in
Fig.~\ref{prewet}. The ionic
strength in the liquid phase is $I=3.9\times10^{-5}$ ($\tilde I=1$mM). $T^*_\text{w}$ is the
transition temperature for first-order
wetting.  Note that
$\mu^*_{0,co}(T^*_{p\text{w},c})\!-\!\mu^*_{0,p\text{w},c}=\tfrac{2/3}{T^*_{p\text{w},c}/T^*_c}\left
[\tfrac { \mu_{0, co }
(T^*_{p\text{w},c})-\mu_{0,p\text{w},c}}{u}\right]$.} 
\label{tab:1}
\end{table}


\section{Conclusions and Summary}\label{CS}

We have investigated wetting of a charged substrate by an
electrolyte solution with
a focus on the influence of the ionic strength $I$ and of the substrate surface charge $\sigma$ on
the wetting
behavior. First, we have investigated a lattice-gas model for the salt-free, i.e., pure solvent
(Fig. \ref{wet})
providing a reference system relative to which the influence of the electrostatic
interaction can be compared.
The results for the salt-free case are in good agreement with previous
studies
\cite{Pandit1983, Pandit1982}. We have calculated the effective interface
potential $\omega(l)$ which facilitates the transparent identification of the order of the
wetting transition (see Fig. \ref{transition}). Depending on the value of the
ratio $u_\text{w}/u$ of the strengths of the substrate potential and of the particle-particle
interaction, the model can exhibit layering transitions when
gas-liquid coexistence is
approached along an isotherm (see Fig. \ref{logsolvent}). In the
next step we have analyzed
quantitatively the effects of the ionic strength and of the surface
charge density on
the order and on the transition temperature of the wetting transition. Concerning the order of the
transition we have found that electrostatic forces induce a first-order wetting
transition, even for very small surface charges (see Fig. \ref{first}). Within our model, for
$\sigma=0$ the
transition is second order and occurs at the same temperature as in the salt-free case. 
For a fixed ionic
strength, the wetting temperature $T^*_\text{w}$ decreases with increasing
surface charge density of
the substrate. This is due to the increasing substrate-fluid attraction as the substrate surface
charge is increased. If systems, which differ only with respect to the ionic strength $I$, are
compared, the one with smaller $I$ has the lower wetting transition temperature
$T^*_{\text w}$ (see Fig. \ref{Tw-sig}). When bulk coexistence is approached along an isotherm, in
the case of a first-order wetting transition, i.e., if $\sigma\neq0$, the model exhibits
first-order layering transitions in addition to prewetting (Fig. \ref{layersandpw}). This leads to
a series of triple points in the surface phase diagram (see Fig. \ref{phase-diagram}). We have also
studied the influence of the surface charge
density on the prewetting lines. We have found
that the
prewetting line becomes shorter as the surface charge density is decreased (Fig. \ref{prewet}).
 
Although our lattice model differs significantly from the continuum models used in Refs.
\cite{Denesyuk2004,
Oleksy2009}, we have arrived at similar conclusions concerning the trend that adding ions
promotes \textit{first}-order wetting transitions. Accordingly, this result can be considered to be
robust.
Within our approach one is able to study \textit{wide} interfacial regions and therefore small
ionic strengths which was not possible within the model studied in Ref. \cite{Oleksy2009}. Our study
also
includes a discussion of prewetting, 
providing a more complete description of the wetting properties of electrolytes.  In agreement
with Refs. \cite{Denesyuk2004, Kayser1988} the growth law of
the film thickness for complete wetting along an isotherm is not changed by adding ions to the
solvent, in spite of their long-ranged Coulombic interaction (Figs. \ref{logsolvent} and
\ref{layersandpw}) .  However, if only counterions are considered, which are donated by the
substrate and the charge of which is opposite to that of the wall, the
film thickness varies as
$l\sim(\mu_{co}-\mu)^{-1/2}$ \cite{Langmuir1938, Kayser1986, Denesyuk2004}.




\end{document}